\RequirePackage[displaymath]{lineno}
\documentclass{appolb}
\usepackage{graphicx}

\newcommand {\rnp}	{$\Delta r_{\rm np}$}
\newcommand {\Nch}	{N_{\rm ch}}
\newcommand {\pt}	{p_{\rm T}}
\newcommand {\mean}[1]	{\langle #1\rangle}
\newcommand {\RuRu}	{$^{96}_{44}$Ru+$^{96}_{44}$Ru}
\newcommand {\ZrZr}	{$^{96}_{40}$Zr+$^{96}_{40}$Zr}
\newcommand {\snn}	{\sqrt{s_{_{\rm NN}}}}
\newcommand {\ebepT}	{[\pt]}
\newcommand{\ac}{{\rm ac}_{2}\{3\}}

\usepackage{color}

\begin{document}
\title{Constraints on neutron skin thickness and nuclear deformations using relativistic heavy-ion collisions from STAR%
\thanks{Presented at Quark Matter 2022}%
}
\author{Haojie Xu (For the STAR collaboration)\address{School of Science, Huzhou University, Huzhou, Zhejiang 313000, China}
}
\maketitle
\begin{abstract}
In these proceedings, we present the measurements of neutron skin thickness and nuclear deformation using isobar \RuRu\ and \ZrZr\ collisions at $\snn=200$ GeV by the STAR detector.  
The significant deviations from unity of the isobar ratios of elliptic flow $v_{2}$, triangular flow $v_{3}$, mean $\pt$ fluctuations $\langle\delta\pt^{2}\rangle/\langle \pt\rangle^{2}$, and asymmetric cumulant $\ac$ indicate large differences in their quadrupole and octuple deformations.
The significant deviations of the isobar ratios of produced hadron multiplicity $\Nch$, mean transverse momentum $\mean{\pt}$, and net charge number $\Delta Q$   indicate a halo-type neutron skin for the Zr nucleus, much thicker than for the Ru nucleus, consistent with nuclear structure calculations. 
We  discuss how we extract the neutron skin thickness, the symmetry energy slope parameter, and deformation parameters from data.
\end{abstract}
  
\section{Introduction}
The isobar collisions, \RuRu\ and \ZrZr, were originally proposed to search for the chiral magnetic effect (CME)~\cite{Voloshin:2010ut}. Based on the  blind analysis with about 2 billion minimum-bias events for each species collected by the STAR experiment, the initial premise of isobar collisions for  CME search  was not realized~\cite{STAR:2021mii}. The CME-related backgrounds were found to differ between isobar collisions~\cite{STAR:2021mii}, suggesting sizeable nuclear structure differences between the two isobar nuclei. Those nuclear structure differences and their consequences in experimental observables had been predicted by energy density functional theory (DFT) calculations~\cite{Xu:2017zcn,Li:2018oec}.
Owing to the large statistics collected by the STAR detector and robust cancellation of systematic uncertainties on the observable ratio 
	$R(X)\equiv \frac{X_{\rm RuRu}}{X_{\rm ZrZr}}$,
isobar collisions provide novel and accurate means to constrain the nuclear structures and strong force parameters. 

Nuclear deformation, a ubiquitous phenomenon for most atomic nuclei, reflects collective motion induced by the interaction between valence nucleons and shell structure.
In most cases, the deformation has a quadrupole shape that is characterized by overall strength $\beta_{2}$ and triaxiality $\gamma$, and/or an octuple shape $\beta_{3}$. 
In relativistic collisions of two nuclei such deformations enhance the fluctuations of bulk observables that are sensitive to initial state geometry~\cite{Zhang:2021kxj}.
The deformation parameters can be constrained from the precision measurements of the ratios between two isobar systems of harmonic anisotropy coefficients $v_{2}$, $v_{3}$,
mean transverse momentum $\ebepT$ fluctuations (mean, variance and skewness), and their Pearson correlation coefficient $\rho(v_{2}^{2}\{2\}, \ebepT)$ ~\cite{Jia:2021oyt}. 
Similar analysis has been performed in Au+Au and U+U collisions~\cite{Jia:2021wbq,Zhang:2022sgk}.

Neutron skin thickness \rnp\ ($\equiv\sqrt{\langle r_{n}^{2} \rangle}-\sqrt{\langle r_{p}^{2}\rangle}$, the root mean square difference between neutron  and proton distributions) of nuclei can infer nuclear symmetry energy.
Such information is of critical importance to the equation of state of dense nuclear matter in neutron stars. 
\rnp\ has traditionally been measured in low-energy hadronic and nuclear scattering experiments.
Recent measurement using parity-violating electroweak interactions by the PREX-II experiment 
has yielded a large neutron skin thickness of Pb nucleus~\cite{PREX-II}, at tension with the world-wide data established in hadronic collisions.
In isobar collisions at relativistic energies, neutron skin was predicted~\cite{Xu:2017zcn,Li:2018oec} to affect event multiplicity and elliptic anisotropy. 
Measurements of those quantities can, in turn, offer an unconventional and perhaps more precise means to probe the neutron skin~\cite{Li:2019kkh}. 
Specifically, the ratios between isobar collisions of the produced hadron multiplicities ($\Nch$)~\cite{Li:2019kkh}, the mean transverse momenta ($\mean{\pt}$)~\cite{Xu:2021kkh}, and the net charge multiplicities ($\Delta Q$)~\cite{Xu:2021qjw} can probe the neutron skin difference between the isobar nuclei. 

\section{Nuclear deformation measurements in isobar collisions}
\begin{figure}[htb]
\centerline{%
	\includegraphics[width=\textwidth]{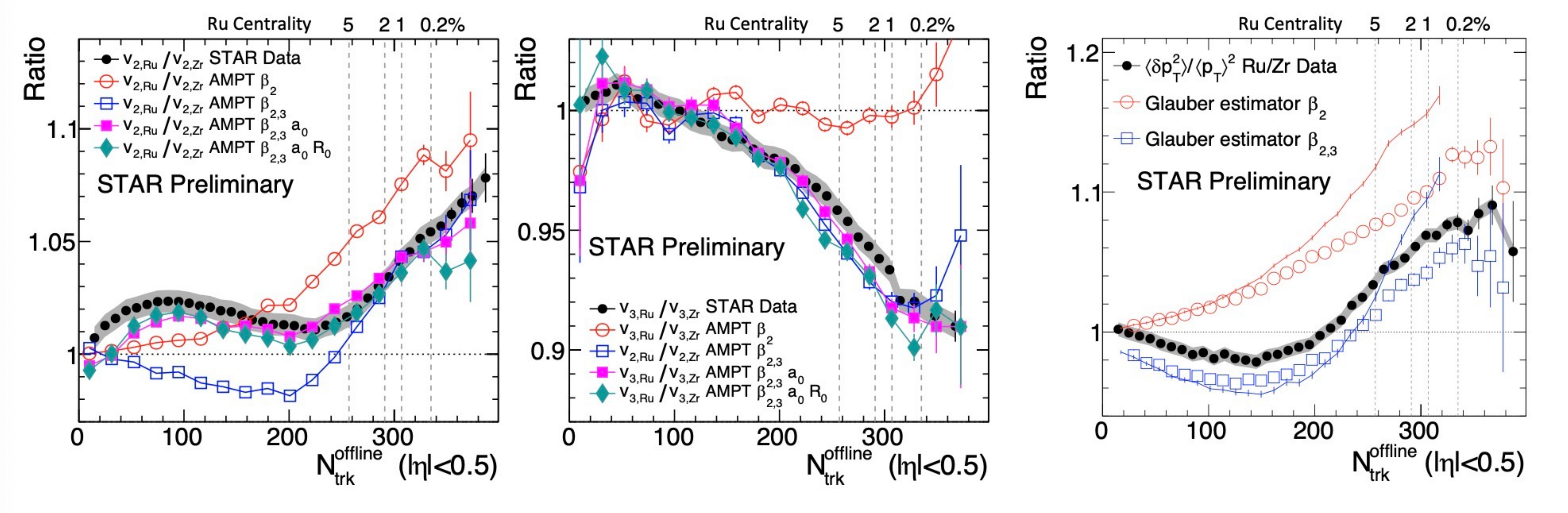}}
\caption{The $R(v_{2})$ (left), $R(v_{3})$ (middle), and $R(\langle \delta\pt^{2}\rangle/\langle \pt\rangle^{2})$ (right) as a function of $N_{\rm trk}^{\rm offline}$ in isobar collisions at $\snn=200$~GeV.}
\label{Fig:1}
\end{figure}

Anisotropic flow in most central collisions is exquisitely sensitive to nuclear deformation. 
The deviations of $R(v_{2})$ and $R(v_{3})$ from unity observed in most central isobar collisions~\cite{STAR:2021mii} indicate a large quadrupole deformation in Ru nucleus and a large octuple deformation in Zr nucleus~\cite{Zhang:2021kxj}. 
Figure~\ref{Fig:1} shows the $R(v_{2})$ and $R(v_{3})$ as a function of charged track multiplicity ($N_{\rm trk}^{\rm offline}$) with $|\eta|<0.5$. 
We simulate events by a multi-phase transport (AMPT) model with varying $\beta_2$ for Ru (and fixed $\beta_2= 0.06$ for Zr~\cite{Pritychenko:2013gwa}) and varying $\beta_3$ for Zr (and fixed $\beta_3=0$ for Ru)  to match the data to extract the best parameter values.
The extracted quadrupole deformation parameter for Ru is $\beta_{2,Ru}=0.16\pm0.02$ and the octuple deformation parameter for Zr is $\beta_{3,Zr}=0.20\pm0.02$, where the quoted uncertainty includes statistical and systematic uncertainties. 
The AMPT results with those deformation parameters are also shown in Fig.~\ref{Fig:1}.
Those deformation parameters extracted from  isobar collisions are consistent with the measurements from traditional nuclear structure experiments~\cite{Pritychenko:2013gwa,Kibedi:2002ube}. We also show in Fig.~\ref{Fig:1} right panel the isobar ratios of mean $\pt$ fluctuations $R(\langle \delta \pt^{2}\rangle/\langle \pt\rangle^{2})$. The trend can be qualitatively described by the Glauber model with the deformation parameters for Ru and Zr.

\begin{figure}[htb]
\begin{minipage}{0.45\textwidth}
\centerline{%
	\includegraphics[width=\textwidth]{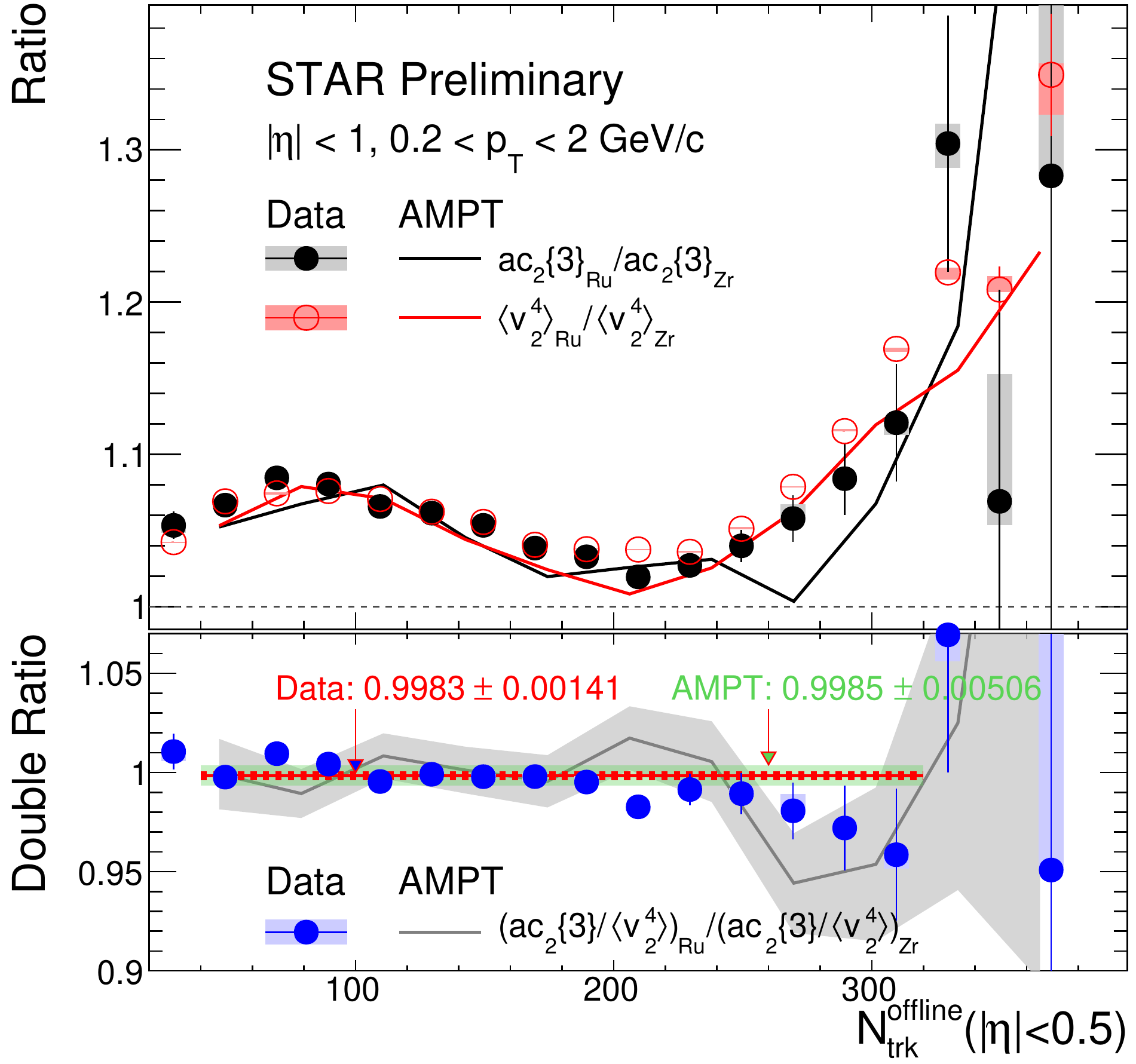}}
\end{minipage}\hfill
\begin{minipage}{0.5\textwidth}
\caption{The three-particle asymmetry cumulant ratio $R(\ac)$ (top) and non-linear response coefficient ratio $R(\chi_{4,22})$ (bottom) as a function of $N_{\rm trk}^{\rm offline}$ in isobar collisions at $\snn=200$~GeV. AMPT simulations are also shown for comparison.}
\label{Fig:2}
\end{minipage}
\end{figure}

The multi-particle correlations are also sensitive to  nuclear deformation. In Fig.~\ref{Fig:2}, we present the  difference between isobar collisions of the three-particle asymmetric cumulant $\ac$. Here
$\ac\equiv\langle\langle e^{i(2\varphi_{1}+2\varphi_{2}-4\varphi_{3}}\rangle\rangle$ is the average of three-particle azimuthal correlations over an ensemble of events. The $R(\ac)$ trend is similar to that of $R(\langle v_{2}^{4}\rangle)$ as shown in Fig.~\ref{Fig:2}. 
The double ratio shown in the bottom panel of Fig.~\ref{Fig:2} indicates that the non-linear response coefficients $\chi_{4,22}=\ac/\langle v_{2}^{4}\rangle$ are almost identical for the two isobar systems. 
Both the trends of $R(\ac)$ and $R(\chi_{4,22})$ can be reproduced by the AMPT model simulations. 
The $R(\ac)$ data, being extra sensitive to flow fluctuations, can help  further constrain the deformation parameters of isobar nuclei~\cite{Zhao:2022uhl}. 

\begin{figure}[htb]
\centerline{%
	\includegraphics[width=6cm]{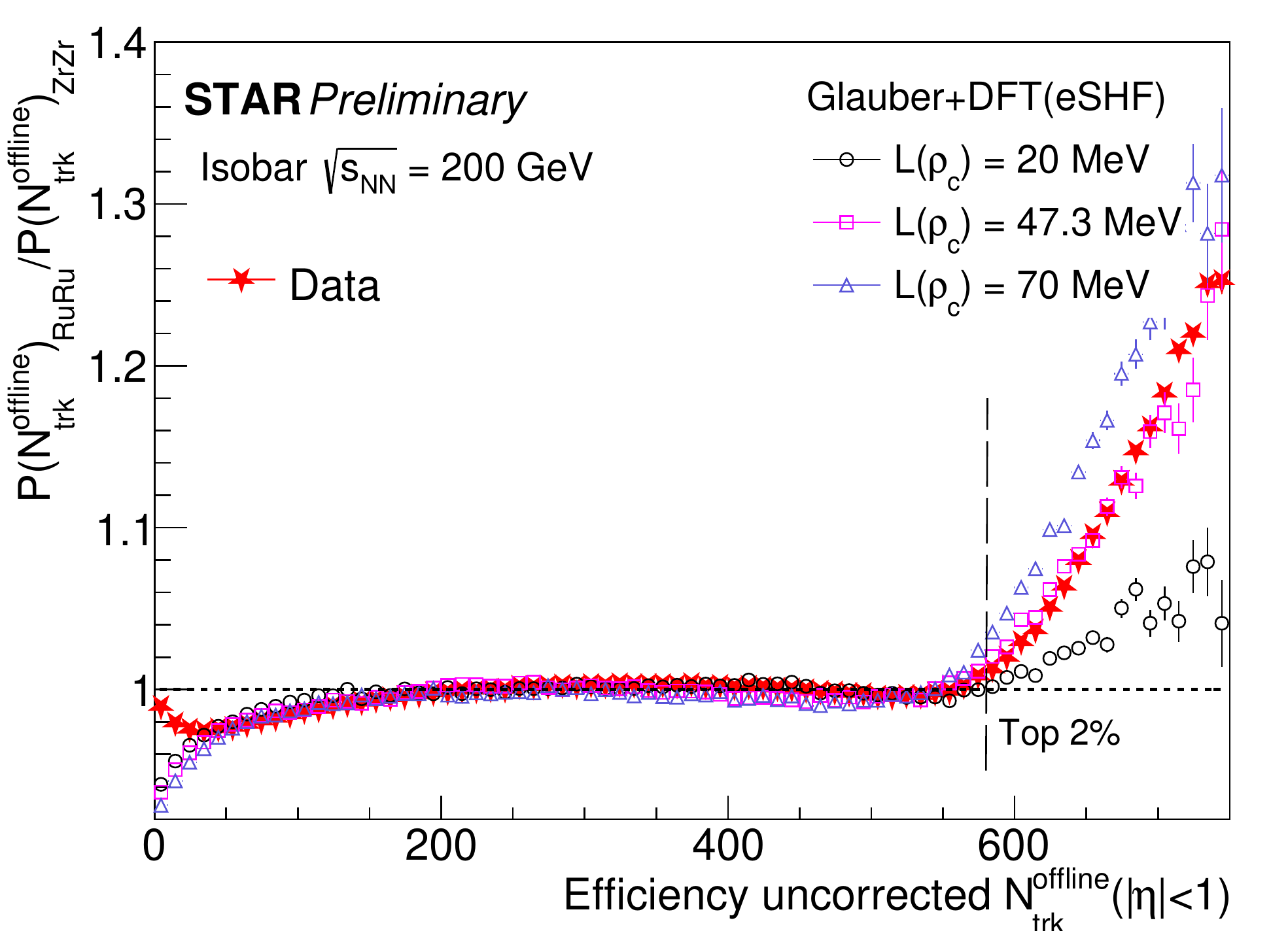}
\includegraphics[width=6cm]{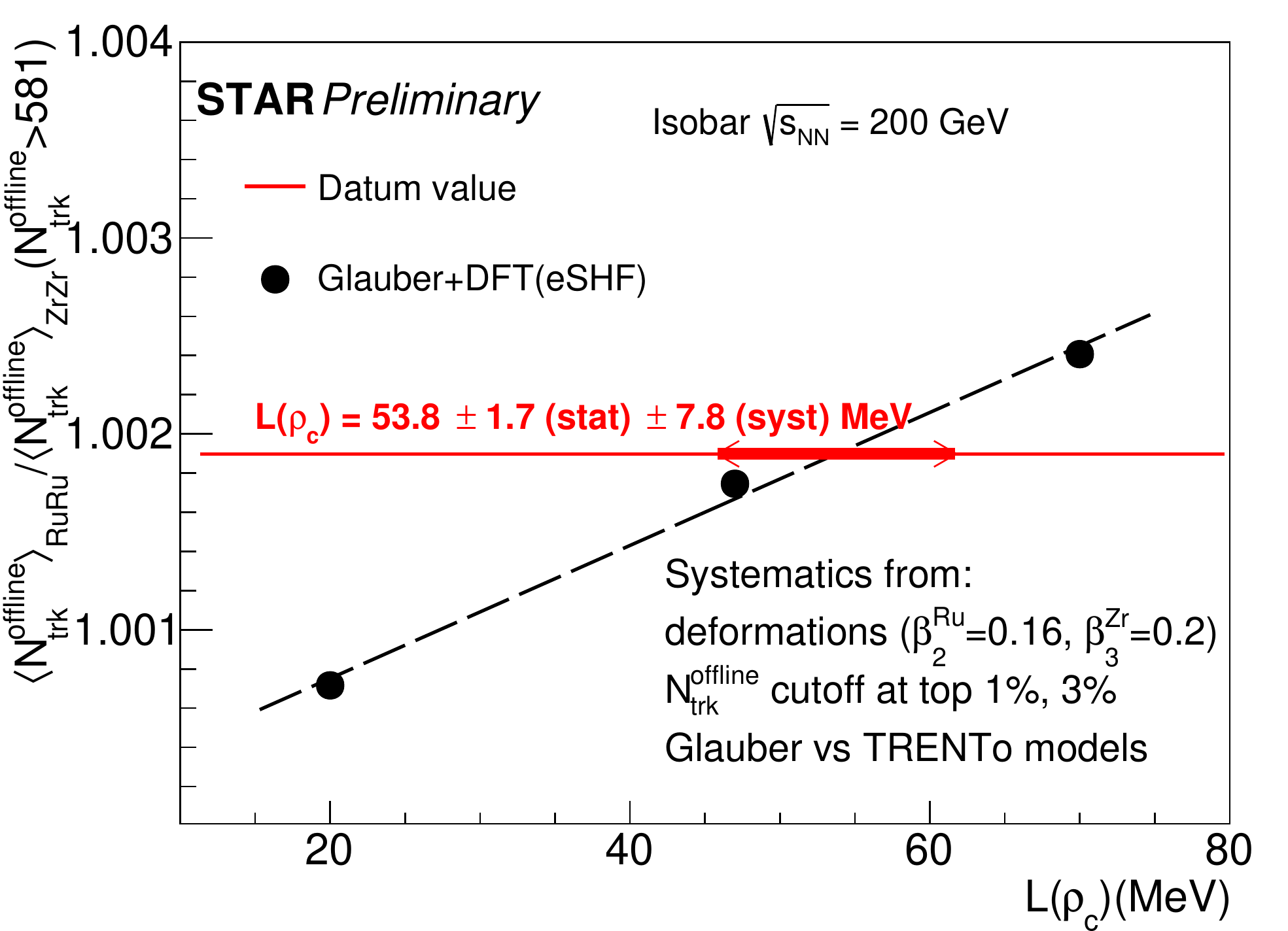}}
\caption{ Left: Ratio of the measured multiplicity ($N_{\rm trk}^{\rm offline}$) distributions in isobar collisions at $\snn=200$ GeV. Also shown are Monte Carlo Glauber model results with nuclear densities calculated by DFT with three $L(\rho_{c})$ values. Right: Ratio of $\mean{N_{\rm trk}^{\rm offline}}$ in  top $2\%$ central collisions as function of $L(\rho_{c})$ from Glauber model. The red line is the measured datum value $1.0019$ with statistical uncertainties smaller than $0.0001$.}
\label{Fig:3}
\end{figure}

\section{Neutron skin measurements}
Nuclear density distributions, and thus the neutron skin thicknesses, depend on the slope parameter $L$ of symmetry energy as a function of nuclear density. 
With a given $L$ of the nuclear interaction potential, the nuclear density can be calculated by the DFT framework.
The event multiplicity produced in heavy-ion collisions is sensitive to the density distributions of the colliding nuclei, and thus the $L$.
In Fig.~\ref{Fig:3}, we present the ratio of the multiplicity distributions measured in isobar collisions within pseudo-rapidity  $|\eta|<1$, and those computed by the Monte Carlo (MC) Glauber model with the  density distributions calculated by DFT with three values of $L(\rho_{c})$~\cite{Li:2019kkh}, where $\rho_{c}=0.11\rho_{0}/0.16\simeq0.11$ fm$^{-3}$ is nuclear subsaturation cross density. The model result with $L(\rho_{c})=47.3$ MeV can reasonably describe the data including the high multiplicity range. In the right panel of Fig.~\ref{Fig:3} we compare the ratios of mean multiplicity at top $2\%$ centrality between data and model calculations 
and obtain $L(\rho_{c}) = 53.8 \pm 1.7 ({\rm stat.}) \pm 7.8 ({\rm sys.})$ MeV. The corresponding neutron skin thicknesses for Ru and Zr are $(\Delta r_{\rm np})_{\rm Ru}=0.051 \pm 0.009$ fm and $(\Delta r_{\rm np})_{\rm Zr}=0.195\pm0.019$ fm, respectively. The systematic uncertainties are estimated with different models (TRENTo vs.~Glauber) and different $N_{\rm trk}^{\rm offline}$ cutoffs, and considering nuclear deformations which are the dominant contribution. 
\begin{figure}[htb]
    \begin{minipage}{0.58\textwidth}
        \centering
	    \includegraphics[width=\textwidth]{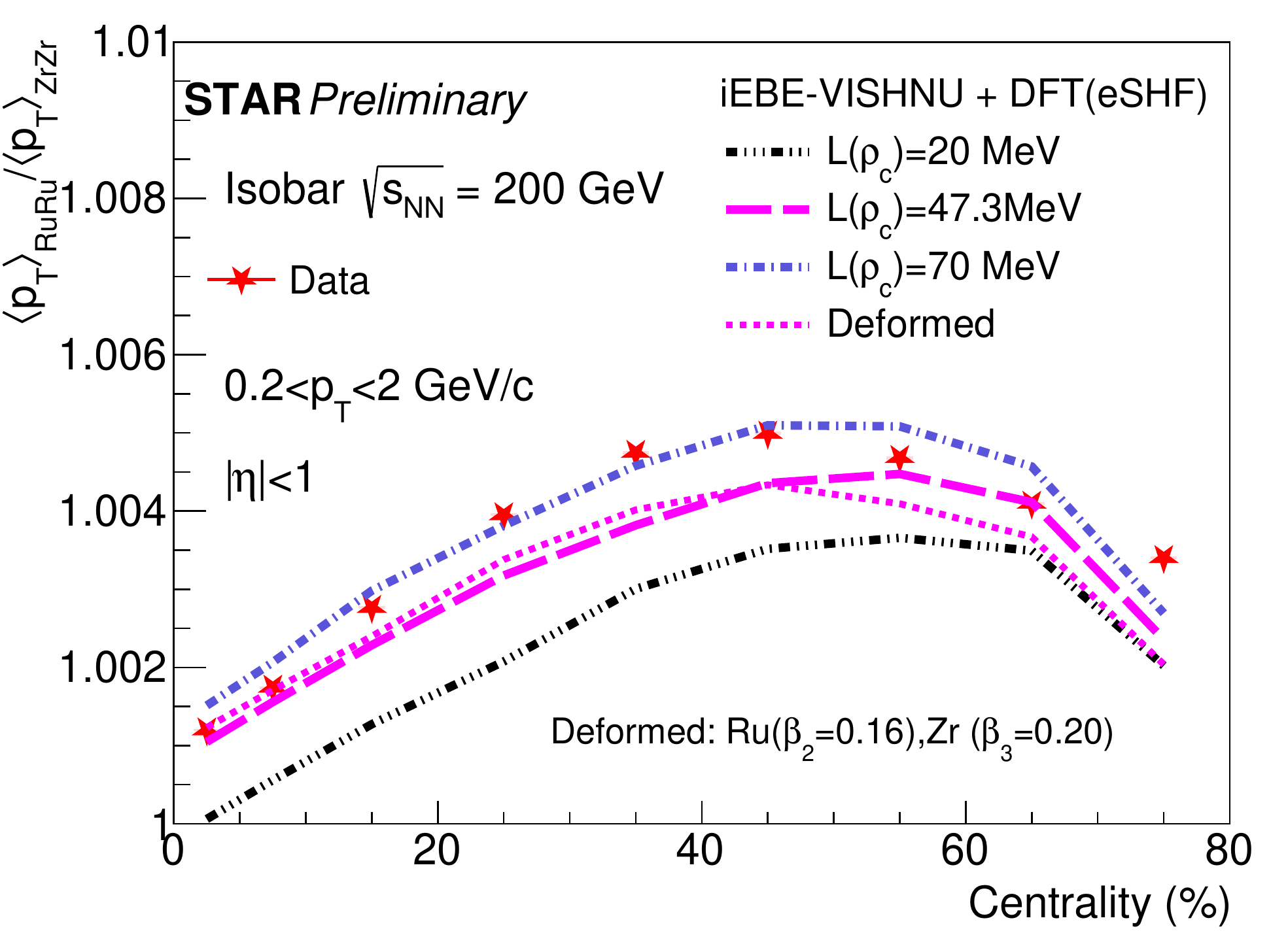}
    \end{minipage}\hfill
    \begin{minipage}{0.40\textwidth}
        \caption{The mean transverse momentum ratio $R(\mean{\pt})$ as a function of centrality in isobar collisions at $\snn=200$ GeV. The three thick curves are obtained from iEBE-VISHNU simulations with three different $L(\rho_{c})$ values, the thin-dashed curve denotes the model simulation with $\beta_{2,\rm Ru}=0.16$ and $\beta_{3,Zr}=0.20$ under $L(\rho_{c})=47.3$ MeV.}
        \label{Fig:4}
    \end{minipage}
\end{figure}

The mean transverse momentum $\mean{\pt}$ also depends on the  size of the colliding nuclei. The sensitivity is the strongest in  most central collisions. This provides  another way to constrain the $L(\rho_{c})$ parameter. Although  $\mean{\pt}$ depends on  bulk properties of the QGP medium, the $R(\mean{\pt})$ in isobar collisions shows weak sensitivity to  shear and bulk viscosities of the QGP medium~\cite{Xu:2021kkh}. The $R(\mean{\pt})$ as a function of centrality is shown in Fig.~\ref{Fig:4}. The trend can be described by  iEBE-VISHNU (Event-By-Event Viscous Israel Stewart Hydrodynamics and UrQMD) model simulations, with the nuclear densities obtained from DFT~\cite{Li:2019kkh,Zhang:2015vaa}. Based on the $R(\mean{\pt})$ values at top $5\%$ centrality, we  extract the slope parameter $L(\rho_{c})=56.8 \pm 0.4 ({\rm stat.}) \pm 10.4 ({\rm sys.})$ MeV, and the corresponding neutron skin thicknesses of the isobar nuclei of $(\Delta r_{\rm np})_{\rm Ru}=0.052\pm0.012$ fm and $(\Delta r_{\rm np})_{\rm Zr}=0.202\pm0.024$ fm. The systematic uncertainties are also dominated by the nuclear deformation effect which can be improved in the future. 
The results extracted from $\mean{\pt}$ are consistent with those from  multiplicity distributions above.

\begin{figure}[htb]
    \begin{minipage}{0.64\textwidth}
        \centerline{%
	    \includegraphics[width=\textwidth]{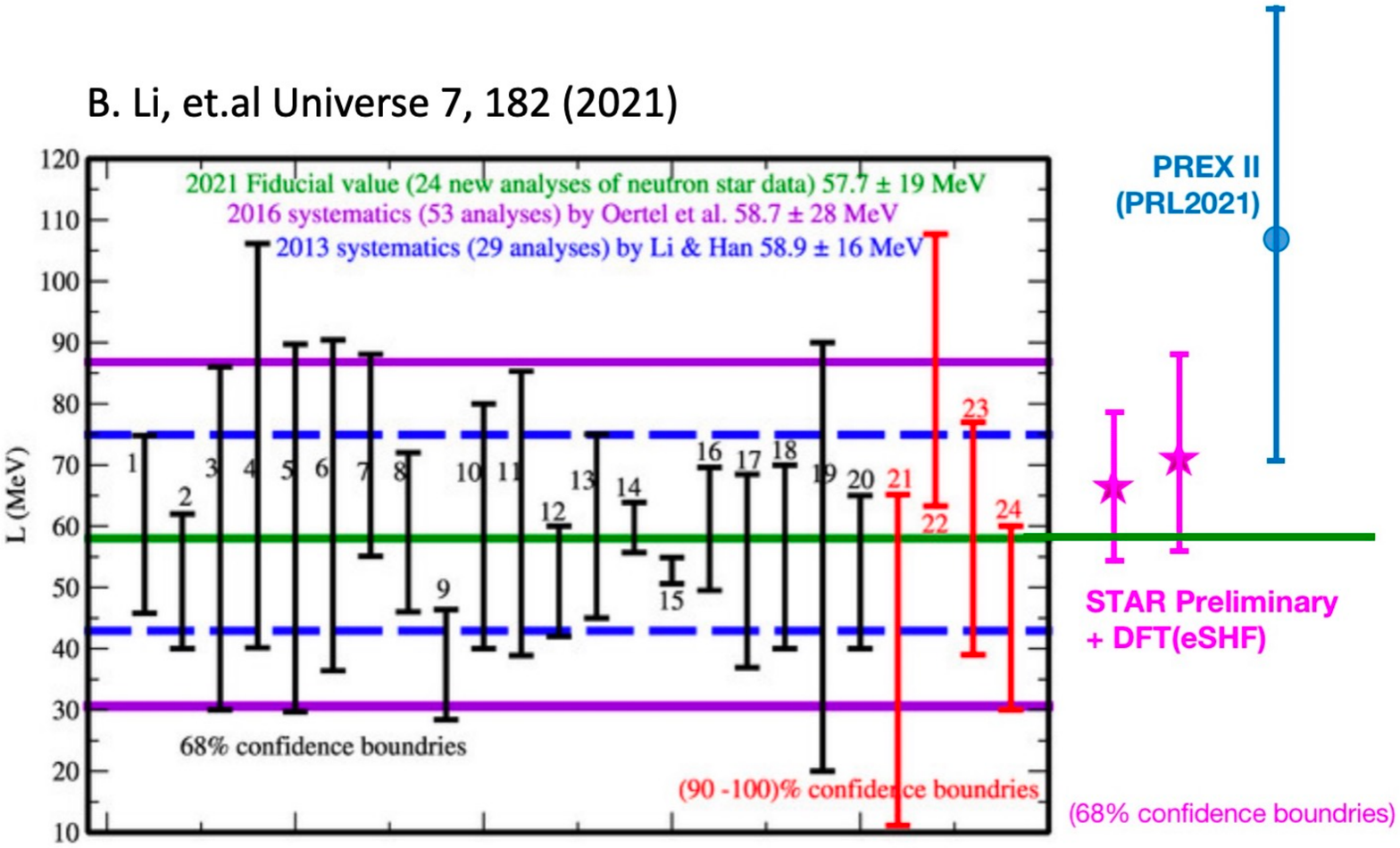}}
    \end{minipage}\hfill
    \begin{minipage}{0.34\textwidth}
        \vspace{8mm}
        \caption{The compilation of world data of $L(\rho)$ from Ref.~\cite{Li:2021thg}. The  $L(\rho_{c})$ values extracted from isobar data are converted into $L(\rho)$ for comparison. The PREX-II data are taken from Ref.~\cite{PREX-II}}
        \label{Fig:5}
    \end{minipage}
\end{figure}

We compare in Fig.~\ref{Fig:5} our $L({\rho})$ (the slope parameter of symmetry energy at saturation density $\rho_{0}$) results with a compilation of world data from traditional nuclear structure experiments~\cite{Li:2021thg}. Our results are consistent with world data with comparable precision. 

\section{Summary}
The isobar \RuRu\ and \ZrZr\ collision data collected by STAR at $\snn=200$ GeV provide novel means to probe the nuclear structure and deformation of the isobar nuclei. From the isobar ratios of the measured anisotropic flow, multiplicity, and mean transverse momentum, with the help of DFT calculations, we have extracted the nuclear deformation parameters $\beta_2, \beta_3$, and neutron skin thicknesses $\Delta r_{\rm np}$ of the isobar nuclei, and the density slope parameter of symmetry energy $L(\rho_{c})$. The results are consistent with world-wide data from traditional nuclear scattering experiments with comparable precision. 
\newline

This work was supported in part by the National Natural Science Foundation of China under Grant Nos. 11905059, 12035006, 12075085.

\end{document}